\documentclass[aps,prd,showpacs,twocolumn,nofootinbib,preprintnumbers]{revtex4-1}

\usepackage{bm}
\usepackage{latexsym}
\usepackage{natbib}
\usepackage{url}
\usepackage{dcolumn}
\usepackage{color}
\usepackage{amsfonts,amssymb,amsmath}
\usepackage{graphicx,epsfig}
\usepackage{psfrag}
\usepackage{subfigure}
\usepackage{natbib}

\begin{document}
\title{The BICEP2 data and  a  single  Higgs-like interacting  tachyonic  field}

\author{Murli Manohar Verma}
\email[]{sunilmmv@yahoo.com}
\author{Shankar Dayal Pathak}
\email[]{prince.pathak19@gmail.com}

\affiliation{Department of Physics, University of Lucknow, \\ Lucknow 226 007, India}

\begin{abstract}
It is proposed  that the recently announced BICEP2 value of tensor-to scalar ratio $r\sim0.2$ can be explained as containing  an extra contribution from the recent acceleration  of the universe. In fact this contribution, being robust, recent  and of much longer duration (by a large order of magnitude) may dominate the contribution  from the inflationary origin. In a possible scenario,   matter (dark or baryonic) and  radiation etc. can emerge from a single Higgs-like tachyonic scalar field in the universe through a physical mechanism not yet fully  known to us. The components  interact among themselves to achieve the thermodynamical equilibrium in the evolution of the universe. The  field potential for the present acceleration of the universe would give a boost to the  amplitude of the tensor fluctuations of gravity waves generated by the early inflation  and the net effects  may be higher than the earlier PLANCK bounds. In the process, the dark energy, as a cosmological constant  decays into creation of  dark matter.   The diagnostics  for the  three-component,  spatially homogeneous  tachyonic scalar field are discussed in detail.  The components of  the field with perturbed equation of state are taken  to interact mutually and the  conservation of energy for individual components  gets violated. We  study mainly the $O_{m}(x)$  diagnostics  with the observed set of $H(z)$ values at various redshifts,   and the dimensionless state-finders  for these interacting components. This analysis provides  a strong case for the interacting dark energy in our model.

\end{abstract}
\maketitle

\section{Introduction}

\label{1}
The recent announcement of BICEP2 observations \cite{a0} of the tensor-to-scalar ratio $r\sim0.20$ claim  the B-mode polarisation signatures to be produced  completely due to the primordial gravity waves arising from the early inflation. However, we think that this belief is  misplaced and the observed  value of $r$ must include the contributions not only from the early inflation but also  the contributions from the present acceleration of the universe. In fact, this acceleration being of the recent origin and of a much longer duration (by many orders) must contribute significantly, and so  the later contributions on the B-mode polarisation of the Cosmic Microwave Background Radiation (CMBR) must be more marked than due to the early inflation, whose sole contribution was reported by Planck \cite{a00}.

We recall that the cosmological and astrophysical observations support the fact our universe is in accelerated expansion phase \cite{a1,a2,a3}, albeit  the exact form of scale factor has not yet  been  fixed  by any observations. Heading  from this motivation previously we adopted the  quasi-exponential expansion that can  also  produce the significant tensor fluctuations of spacetime \cite{a4}.
In \cite{q2} the cosmological constant $\Lambda$ with the energy density of the self-interaction of scalar bosons bound in a condensate is identified. In this approach, $\Lambda$ decay provides with a dynamical realization of spontaneous symmetry breaking.  The detailed  kinematics of $\Lambda$ decay and the back reaction of the  decay products on the $\Lambda$ dynamics are given in \cite{q3}. The slow evaporation regime is found here for a wide range of possible parameters of particle interactions.

The present  paper is organised into five sections. In section $(2)$ we study the mutual interaction among  the Higgs-like  tachyon field components wherein the  gravity waves unleashed by the earlier inflation may be further boosted by the present  acceleration of the universe, and may appear beyond the PLANCK upper bounds \cite{a00}. Section $(3)$ is laid  for $O_{m}(x)$-diagnostic and section   $(4)$  is  devoted towards $O_{m}3(x)$-diagnostic and statefinder parameters  for interacting components of tachyonic field. The Lagrangian for the tachyonic scalar field appears from string theory \cite{n1} and is  given  from the action

\begin{eqnarray}\mathcal{A}=\int d^{4}x \sqrt{-g} \left(\frac{R}{16\pi G}- V(\phi)\sqrt{1-\partial^{i}\phi\partial_{i}\phi}\right)\label{n1}\end{eqnarray} as
\begin{eqnarray}\mathcal{L}=-V(\phi)\sqrt{1-\partial^{i}\phi\partial_{i}\phi}\label{n2}\end{eqnarray}
 whereas  the energy-momentum tensor

\begin{eqnarray}T^{ik}=\frac{\partial\mathcal{L}}{\partial(\partial_{i}\phi)}\partial^{k}\phi-g^{ik}\mathcal{L}\label{n3}\end{eqnarray} leads to  pressure and
energy density of the tachyonic scalar field as
\begin{eqnarray}P=- V(\phi)\sqrt{1-\partial^{i}\phi\partial_{i}\phi}\label{n4}\end{eqnarray}  and \begin{eqnarray}\rho=\frac{V(\phi)}{\sqrt{1-\partial^{i}\phi\partial_{i}\phi}}.\label{n5}\end{eqnarray}
For spatially homogeneous tachyonic scalar field we have
\begin{eqnarray}P=- V(\phi)\sqrt{1-\dot{\phi}^{2}}\label{n6}\end{eqnarray}
\begin{eqnarray}\rho=\frac{V(\phi)}{\sqrt{1-\dot{\phi}^{2}}}\label{n7}.\end{eqnarray}
Here,  we assume  that radiation with equation of state \[w_{r}=1/3\]  also exists as  one inherent  component of  same tachyonic scalar field. Due to some physical mechanism  not known in detail at present to us, but that may be like a Higgs  mechanism,  we can split  the expressions  (\ref{n6}) and  (\ref{n7}) as
\begin{eqnarray}P=-\frac{V(\phi)}{\sqrt{1-\dot{\phi}^{2}}} + \frac{\dot{\phi}^{2}V(\phi)}{\sqrt{1-\dot{\phi}^{2}}} + 0 \label{n8}\end{eqnarray}

\begin{eqnarray}\rho=\frac{V(\phi)}{\sqrt{1-\dot{\phi}^{2}}} + \frac{3\dot{\phi}^{2}V(\phi)}{\sqrt{1-\dot{\phi}^{2}}} - \frac{3\dot{\phi}^{2}V(\phi)}{\sqrt{1-\dot{\phi}^{2}}}.\label{n9}\end{eqnarray}
From (\ref{n8}) and  (\ref{n9}) it is seen  that when we include  radiation in tachyonic scalar field then one new exotic component also appears (say,  exotic matter since its energy density is negative) with zero pressure. Thus,  the tachyonic scalar field resolves into  three components say $a$, $b$ and $c$. The pressure and energy density  of $a$  is given as \[P_{a}=-\frac{V(\phi)}{\sqrt{1-\dot{\phi}^{2}}}, \quad \rho_{a}=\frac{V(\phi)}{\sqrt{1-\dot{\phi}^{2}}} \Rightarrow w_{a} = -1 = w_{\lambda}.\] This is nothing but the  `true' cosmological constant because of its equation of state  being $ w_{\lambda}=-1$.

The second component with   \[P_{b}=\frac{\dot{\phi}^{2}V(\phi)}{\sqrt{1-\dot{\phi}^{2}}},  \quad \rho_{b}=\frac{3\dot{\phi}^{2}V(\phi)}{\sqrt{1-\dot{\phi}^{2}}} \Rightarrow w_{b} = 1/3 \] can be identified as  radiation with \[w_{r}=1/3.\] The  last component is characterised by  \[P_{c}=0,\quad \rho_{c}= -\frac{3\dot{\phi}^{2}V(\phi)}{\sqrt{1-\dot{\phi}^{2}}} \Rightarrow w_{c}=0.\] This component  mimics dust matter but has  negative energy density. The exotic matter \footnote{The anti-particles may be interpreted to  have negative energy density, and so, the  anti-Dirac fermions possess the  negative energy density in contrast to their  particles. Since the  Majorana fermions are self-anti-particles,  their negative energy state changes  into the one with  positive energy  and vice-versa. Therefore, it may be plausible that this exotic matter exists in  such incarnation of dust. The other possible   alternatives  for explanation of this negative energy density of the exotic matter may indicate some new physics beyond the Standard Model of particle physics.} may  include  the Dirac fermions as well as  the Majorana fermions whence the negative energy states turn into the positive energy states\cite{a5,b1}.

In our earlier work \cite{q1} we  allowed a  small time dependent perturbation in  the equation of state(EoS) of the cosmological constant with \[\bar{w}_{\lambda}=-1+\varepsilon(t).\]

Thus, with  the perturbed EoS, the true cosmological constant becomes  a shifted cosmological parameter.  This has a bearing upon the  EoS of radiation and exotic matter, both. Therefore,  these two entities  turn into  shifted radiation and shifted exotic matter respectively.  With fixed energy density of field components, the expressions for the  energy density and  pressure of each component are  given as below. For the shifted cosmological constant one has

\begin{eqnarray}
    \bar{\rho}_{\lambda}=\frac{V(\phi)}{\sqrt{1-\dot{\phi}^{2}}}\label{n10}
\end{eqnarray}

\begin{eqnarray}
    \bar{p}_{\lambda}=\frac{-V(\phi)}{\sqrt{1-\dot{\phi}^{2}}}+\frac{\varepsilon V(\phi)}{\sqrt{1-\dot{\phi}^{2}}}\label{n11}
\end{eqnarray}
and  \[\bar{w}_{\lambda}=-1+\varepsilon(t).\]

For shifted radiation, we have

\begin{eqnarray}
    \bar{\rho}_{r}=\frac{3\dot{\phi}^{2}V(\phi)}{\sqrt{1-\dot{\phi}^{2}}}\label{n12}
\end{eqnarray}
\begin{eqnarray}
    \bar{p}_{r}=\frac{(1+3\varepsilon)\dot{\phi}^{2}V(\phi)}{\sqrt{1-\dot{\phi}^{2}}}\label{n13}
\end{eqnarray} with \[\bar{w}_{r}=\frac{1}{3}+\varepsilon .\]

In presence of perturbation the zero pressure of exotic matter turns into negative non-zero pressure due to  shifted exotic matter which would  also accelerate the universe like dark energy. Thus,  the  energy density and pressure for shifted exotic matter are now,  respectively,   given as

\begin{eqnarray}
    \bar{\rho}_{m}=\frac{-3\dot{\phi}^{2}V(\phi)}{\sqrt{1-\dot{\phi}^{2}}}\label{n14}
\end{eqnarray}

\begin{eqnarray}
  \bar{p}_{m}=p_{\phi}-\bar{p}_{\lambda}-\bar{p}_{r}=\frac{-\varepsilon(1+3\dot{\phi}^{2})V(\phi)}{\sqrt{1-\dot{\phi}^{2}}}\label{n15}
\end{eqnarray} with \[\bar{w}_{m}=\frac{\varepsilon(1+3\dot{\phi}^{2})}{3\dot{\phi}^{2}}.\]

\section{Interaction among the components of tachyonic scalar field}

The interacting dark energy models have been recently proposed by several authors \cite{x1,x2,x3,x4,x5,x6}. Why  must the components of tachyonic scalar field interact? This is  one of the interesting questions  about interaction. Since all components are in thermodynamic non-equilibrium,  therefore,  to achieve  an  equilibrium state they must fall into  mutual  interaction. We propose  that the currently  ongoing acceleration must also  boost the tensor-scalar  fluctuations caused by  the early inflation, and the interaction among the components may be responsible for this boost beyond the Planck bounds \cite{a00}.   With this motivation we study the interaction of these components   assuming that even though  the total energy of the  perturbed field (spatially homogeneous) is kept  conserved,   yet  during interaction it can  get  reasonably  violated for individual components. The equations for conservation of energy with interaction  are  given as

\begin{eqnarray}
\dot{\bar{\rho}}_{\lambda}+3H(1+\bar{w}_{\lambda})\bar{\rho}_{\lambda}=-Q_{1}\label{n16}
\end{eqnarray}

\begin{eqnarray}
\dot{\bar{\rho}}_{r}+3H(1+\bar{w}_{r})\bar{\rho}_{r}=Q_{2}\label{n17}
\end{eqnarray}

\begin{eqnarray}
\dot{\bar{\rho}}_{m}+3H(1+\bar{w}_{m})\bar{\rho}_{m}=Q_{1}-Q_{2}\label{n18}
\end{eqnarray}
where $Q_{1}$ and  $ Q_{2} $ are the interaction strengths. In the above expressions (\ref{n16}),(\ref{n17}) and (\ref{n18}) the following  broad conditions must  govern the dynamics.

\textbf{Condition(I)}  $\mid Q_{1}\mid > \mid Q_{2}\mid$.  This corresponds  to  the  following cases,

\begin{description}
  \item[(i)] If $Q_{1} > 0$,  $Q_{2} > 0$  then the right hand side of (\ref{n16}) is negative while (\ref{n17}) and (\ref{n18}) are positive, respectively. This means that there is  energy transfer from shifted cosmological parameter to shifted radiation and shifted exotic matter,  respectively. Thermodynamics allows this kind of transfer of energy.
  \item[(ii)]  $Q_{1} < 0$,  $Q_{2} < 0$ implies  that there is an  energy transfer to shifted cosmological parameter from shifted radiation and shifted exotic matter.
\end{description}

\textbf{Condition(II)}  $ \mid Q_{2}\mid > \mid Q_{1}\mid$.

\begin{description}
  \item[(i)]  $Q_{2} > 0$,  $Q_{1} > 0$  would make  the right hand side of  (\ref{n17}) as  positive and  (\ref{n16}) and (\ref{n18}) as  negative. This shows  that there is an  energy transfer to shifted radiation from shifted cosmological parameter  and shifted exotic matter.  Thermodynamics again does  not allow this kind interaction.
  \item[(ii)]  $Q_{1} < 0$,  $Q_{2} < 0$ makes way for the   energy transfer from shifted radiation to  shifted exotic matter and  shifted cosmological parameter.
\end{description}

\textbf{Condition(III)} If  $Q_{2} = Q_{1}=Q$   then  we have the following possibility

\begin{description}
  \item[(i)] $Q>0$ leads to an  energy transfer to shifted radiation from shifted cosmological parameter, while the  shifted exotic matter remains free from interaction with its  energy density held conserved. This type of interaction holds compatibility with  the laws of thermodynamics.
  \item[(ii)] If  $Q<0$, energy would flow  from shifted radiation to shifted cosmological parameter, whereas the  shifted exotic matter does  not get involved  in interaction mechanism.  Thus,  the conservation of energy for shifted exotic matter holds good.

      \item[(iii)] As an  alternative, $Q=0$  would pull   the components of tachyonic scalar field  out of mutual interaction  like the standard $\Lambda$CDM model.
\end{description}
 The second case of condition (I) and condition(II) violate the laws of thermodynamics,  therefore,  we are  not interested  in these types of interactions. The interaction of type condition (iii) has been previously discussed for two components of  tachyonic scalar field in our earlier work \cite{a4}.

The positivity of the  quantity $Q_{1}-Q_{2}$   implies that $Q_{1}$  should be large and positive. For if  $Q_{1}$ had been  large and negative then the second law of thermodynamics would have been violated and  the cosmological constant (as the dark energy candidate) would have  dominated much  earlier withholding  the   structure formation against the present observations. Also,  $Q_{2}$ should be positive and small since  if it is negative and large then conservation of energy of tachyonic field is violated. The interaction strengths $Q_{1}$ and $Q_{2}$ should  depend on temperature also, but  due to mathematical simplification following the Occam's razor,  we consider the interaction strength as  independent of temperature. On the left hand side of energy conservation equation are Hubble parameter (reciprocal of dimension of time) and energy density thus it is natural choice that the interaction strength should be function of Hubble parameter and energy density.
Several authors have proposed different forms of $Q$ \cite{y1,z1,z2,z3}. Owing  to the lack of information regarding the  exact nature of dark matter and dark energy (as the cosmological constant or else) we cannot yet  fix the exact form of interaction strength.  Thus,  with  this motivation  we present the form of interaction term  heuristically  as  function of time rate of change in  energy densities as

                                 \begin{enumerate}
                                   \item $Q_{1}=\alpha\dot{\bar{\rho_{\lambda}}}$
                                   \item $Q_{2}=\beta\dot{\bar{\rho_{r}}}$
                                 \end{enumerate}

                                where $\alpha, \beta$ are proportionality constant. The total conservation of energy of field is ensured by

\begin{eqnarray}
\dot{\bar{\rho}}_{\phi}+3H(1+\bar{w}_{\phi})\bar{\rho}_{\phi}=0\label{n20}
\end{eqnarray} with $\bar{w}_{\phi}=\dot{\phi}^{2}-1$ and Hubble parameter $H$ given from

\begin{eqnarray}
H^{2}(t)=\frac{8\pi G}{3}[\bar{\rho}_{\lambda}+\bar{\rho}_{r}+\bar{\rho}_{m}]\label{n21}.
\end{eqnarray}

From  (\ref{n16}), (\ref{n17}) and (\ref{n18}),  the functional form of energy density with redshift $z$ is given as
\begin{eqnarray}
\bar{\rho}_{\lambda}=\bar{\rho}^{0}_{\lambda}x^{3\varepsilon/1+\alpha}\label{n22}
\end{eqnarray} where \[\frac{a_{0}}{a}=1+z=x\]

\begin{eqnarray}
\bar{\rho}_{r}=\bar{\rho}^{0}_{r}x^{4+3\varepsilon/1-\beta}\label{n23}
\end{eqnarray} and

\begin{eqnarray}
\bar{\rho}_{m}=\bar{\rho}^{0}_{m}x^{\eta}+\left(\frac{3\varepsilon\alpha\bar{\rho}^{0}_{\lambda}}{3\varepsilon-\eta-\eta\alpha}\right)[x^{3\varepsilon/1+\alpha}-x^{\eta}]\nonumber\\-\left(\frac{\beta\bar{\rho}^{0}_{r}(4+3\varepsilon)}{4+3\varepsilon-\eta+\eta\beta}\right)[x^{4+3\varepsilon/1-\beta}-x^{\eta}]\label{n24}
\end{eqnarray}
where $\eta$ (constant) is  defined as

 \begin{eqnarray}
\eta=\frac{3\dot{\phi}^{2}(1+\varepsilon) + \varepsilon}{\dot{\phi}^{2}}.\label{n25}
\end{eqnarray}
It is clearly seen  that in the absence of perturbation,  $\alpha$  does not play any role in the evolution of the matter component and so delinks it from the coextensive evolution of the shifted cosmological parameter. In that case,  we have \[\varepsilon=0 \Rightarrow \eta=3.\] Then,   (\ref{n24})  would turn into

\[\bar{\rho}_{m}=\bar{\rho}^{0}_{m}x^{3}-\left(\frac{4\beta\bar{\rho}^{0}_{r}}{1+3\beta}\right)[x^{4/1-\beta}-x^{3}]\]
which,  in the absence of interaction($\beta=0$), further yields
\[\bar{\rho}_{m}=\bar{\rho}^{0}_{m}x^{3}\]
as expected in $\Lambda$CDM model.

\begin{figure}[h]
\centering  \begin{center} \end{center}
\includegraphics[width=0.50\textwidth,origin=c,angle=0]{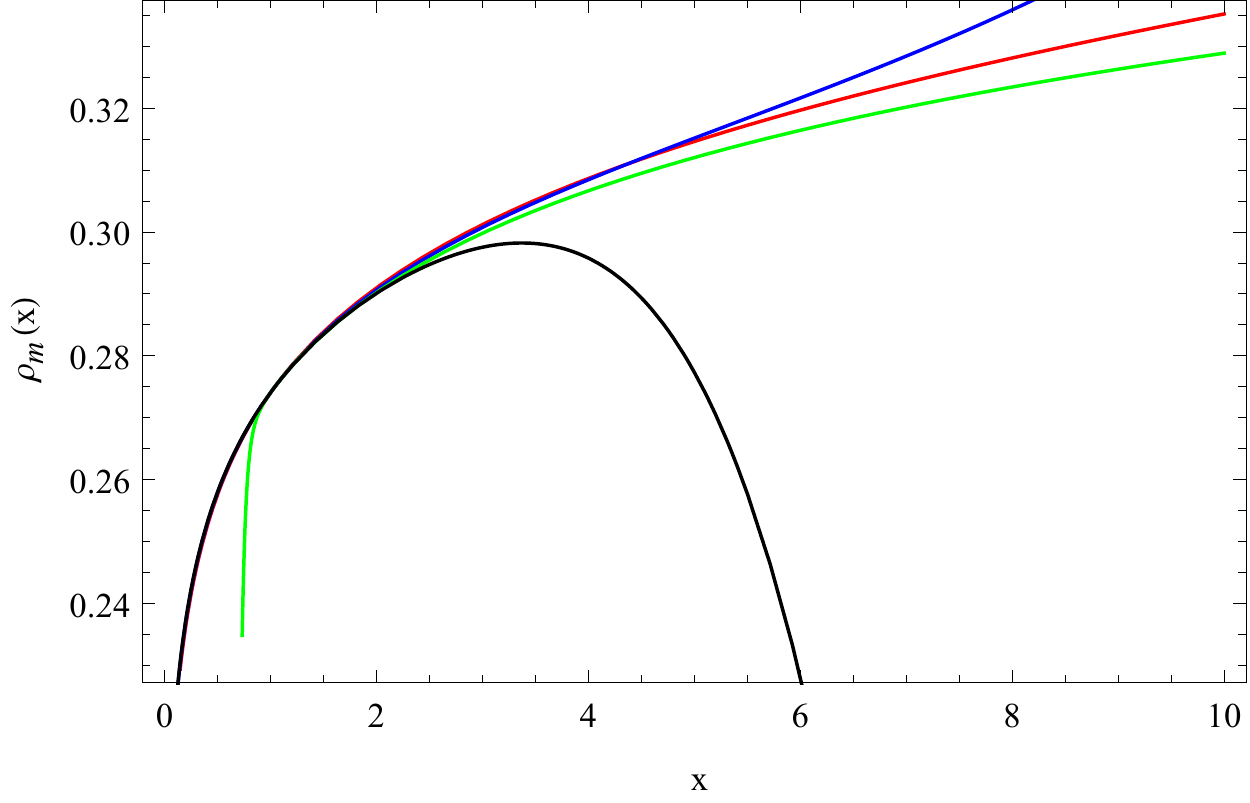}
\caption{\label{fig:i}Plot for variation of shifted exotic matter energy density $\rho_{m}(x)$ with redshift ($x$ for range $0$ to $10$). We assume $\eta=0.001$, $\varepsilon = 0.02$, $\alpha=1.2$ and $\beta=1.2, -1.2, 0.2, -0.2$ for  $\bar{\Omega}^{0}_{m}\simeq 0.274$ and $\bar{\Omega}^{0}_{\lambda}\simeq 0.725$. The red, green, blue and black  curves  correspond to $\beta=1.2, -1.2, 0.2$ and $-0.2$,  respectively.}
\end{figure}

\begin{figure}[h]
\centering  \begin{center} \end{center}
\includegraphics[width=0.50\textwidth,origin=c,angle=0]{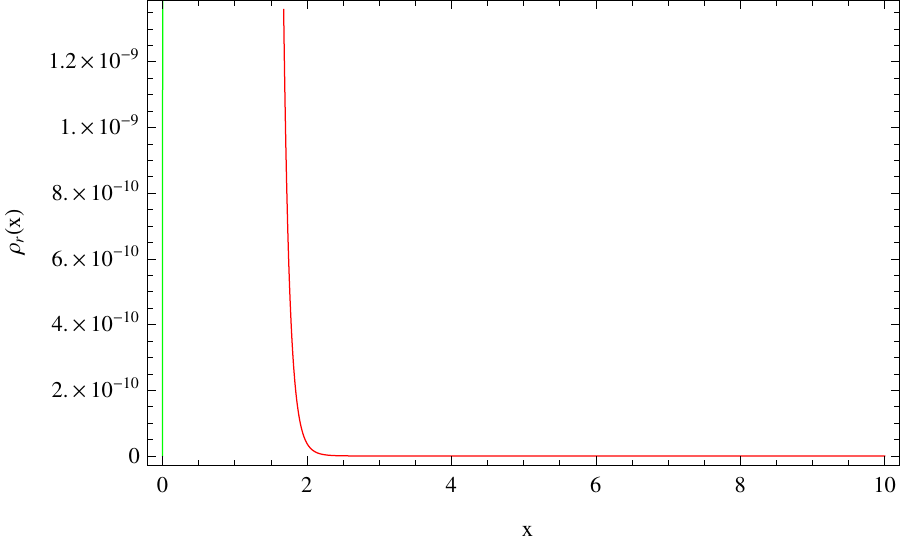}
\caption{\label{fig:i}Variation of shifted radiation energy density $\rho_{r}(x)$ with redshift ($x$ for range $0$ to $10$). We assume $\eta=0.001$, $\varepsilon = 0.02$, $\alpha=1.2$ and $\beta=1.2$ and $-1.2$ for $\bar{\Omega}^{0}_{r}\simeq 0.00005$. The red and green curves  stand for $\beta=1.2$ and $-1.2$ respectively. }
\end{figure}

\begin{figure}[h]
\centering  \begin{center} \end{center}
\includegraphics[width=0.50\textwidth,origin=c,angle=0]{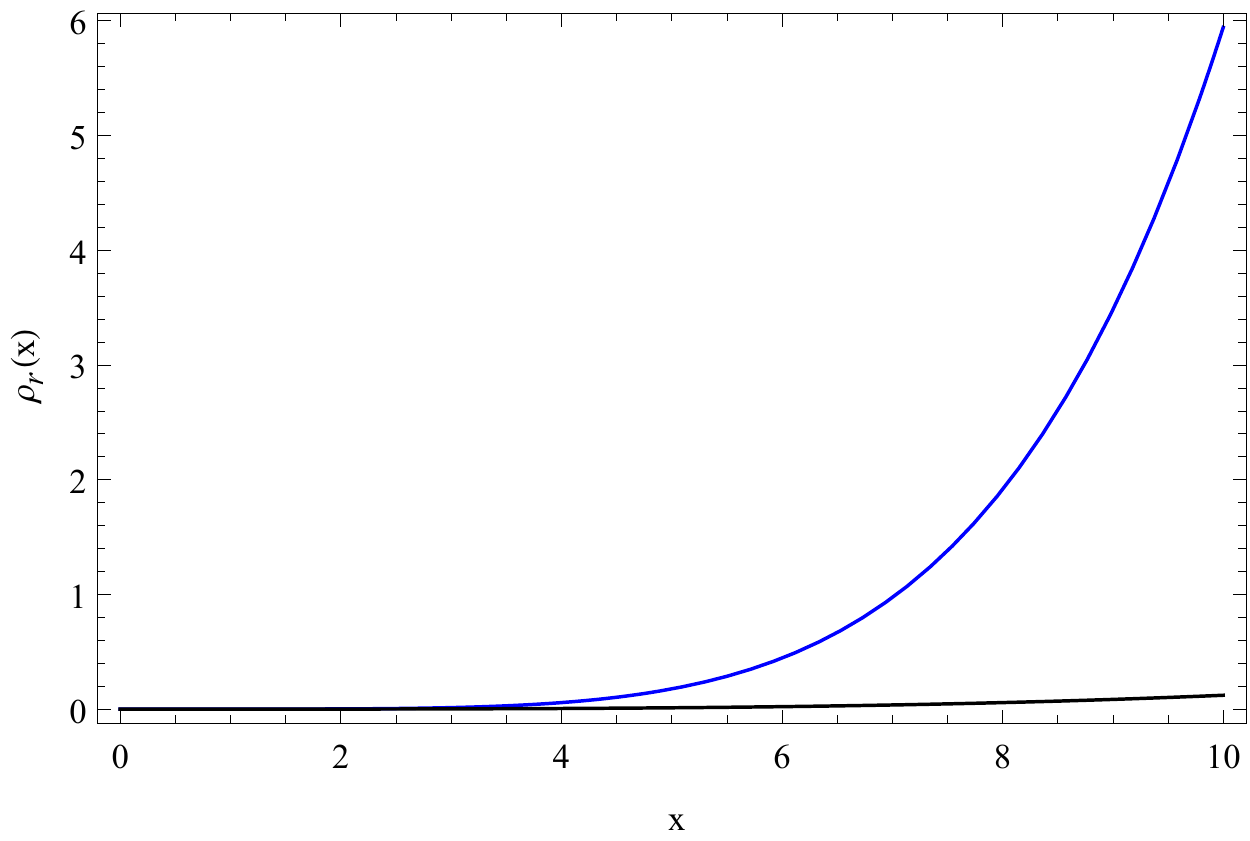}
\caption{\label{fig:i}Variation of shifted radiation energy density $\rho_{r}(x)$ with redshift ($x$ for range $0$ to $10$). We assume $\eta=0.001$, $\varepsilon = 0.02$, $\alpha=1.2$ and $\beta=0.2$ and $-0.2$ for $\bar{\Omega}^{0}_{r}\simeq 0.00005$. The blue and black  curves  refer to  $\beta=0.2$ and $-0.2$,  respectively. }
\end{figure}

\begin{figure}[h]
\centering  \begin{center} \end{center}
\includegraphics[width=0.50\textwidth,origin=c,angle=0]{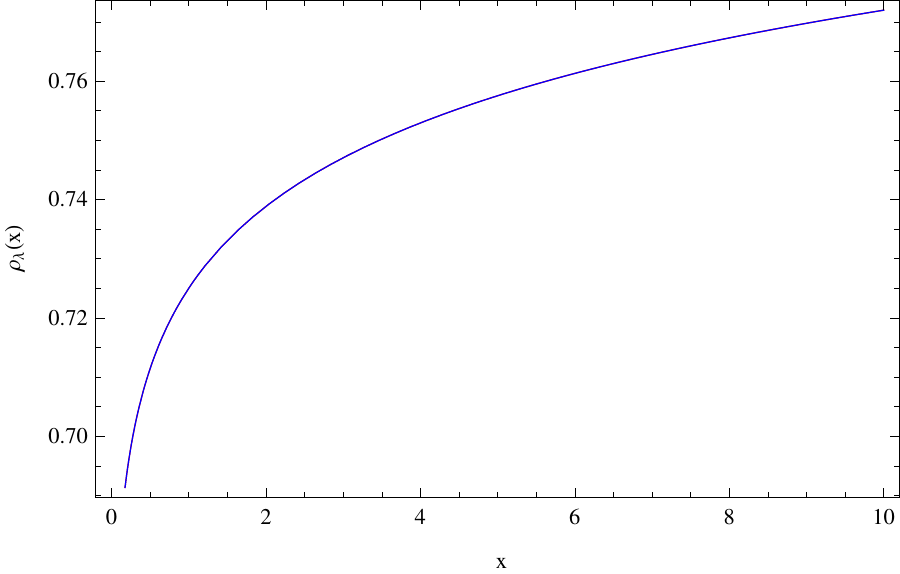}
\caption{\label{fig:i}Variation of energy density for shifted cosmological parameter $\rho_{\lambda}(x)$ with redshift ($x$ for range $0$ to $10$). We assume $\eta=0.001$, $\varepsilon = 0.02$, $\alpha=1.2$  for  $\bar{\Omega}^{0}_{\lambda}\simeq 0.725$. The plot is independent of the chosen values of $\beta$ }
\end{figure}


\section{$O_{m}(x)$ diagnostics for interaction}
\label{sec:3}

Recently, Sahni \emph{et al} \cite{a6} introduced the redshift dependent function \begin{eqnarray}O_{m}(x)=\frac{E^{2}(x)-1}{x^{3}-1}\label{n27}\end{eqnarray} where $E(x)=H(x)/H_{0}$ is the normalized Hubble function

\begin{eqnarray}
E^{2}(x)=(1+M)\bar{\Omega}^{0}_{\lambda}x^{3\varepsilon/1+\alpha}+(1-N)\bar{\Omega}^{0}_{r}x^{4+3\varepsilon/1-\beta}\nonumber\\+(\bar{\Omega}^{0}_{m}-M\bar{\Omega}^{0}_{\lambda}+N\bar{\Omega}^{0}_{r})x^{\eta}\label{n28}
\end{eqnarray} where  the constants  $M$  and  $N$  are respectively   given as

\begin{eqnarray}
M=\frac{3\varepsilon\alpha}{3\varepsilon-\eta-\eta\alpha}\label{n29}
\end{eqnarray}

\begin{eqnarray}
N=\frac{\beta(4+3\varepsilon)}{4+3\varepsilon-\eta+\eta\beta}\label{n30}
\end{eqnarray}  and  $\bar{\Omega}^{0}$  is  the present density parameter in  the  spatially flat $(k=0)$  universe.

\begin{figure}[h]
\centering  \begin{center} \end{center}
\includegraphics[width=0.50\textwidth,origin=c,angle=0]{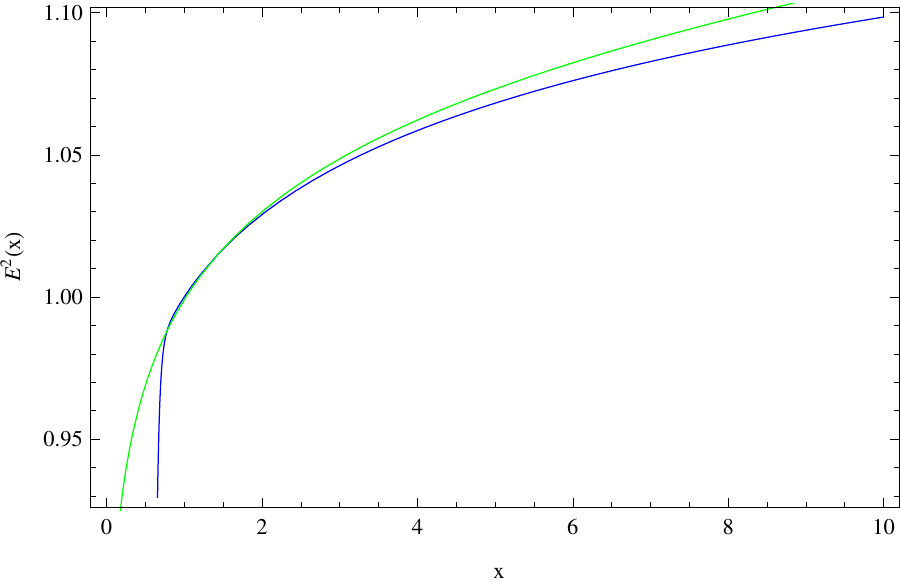}
\caption{\label{fig:i}Normalized Hubble parameter  $E^{2}(x)$ with redshift ($x$ for range $0$ to $10$). We assume $\eta=0.001$, $\varepsilon = 0.02$, $\alpha=1.2$ and $\beta=1.2 $ and $-1.2$ for  $\bar{\Omega}^{0}_{m}\simeq 0.274$ and $\bar{\Omega}^{0}_{\lambda}\simeq 0.725$. Blue and green curves  correspond  to $\beta=1.2$ and $ -1.2 $,  respectively.}
\end{figure}

\begin{figure}[h]
\centering  \begin{center} \end{center}
\includegraphics[width=0.50\textwidth,origin=c,angle=0]{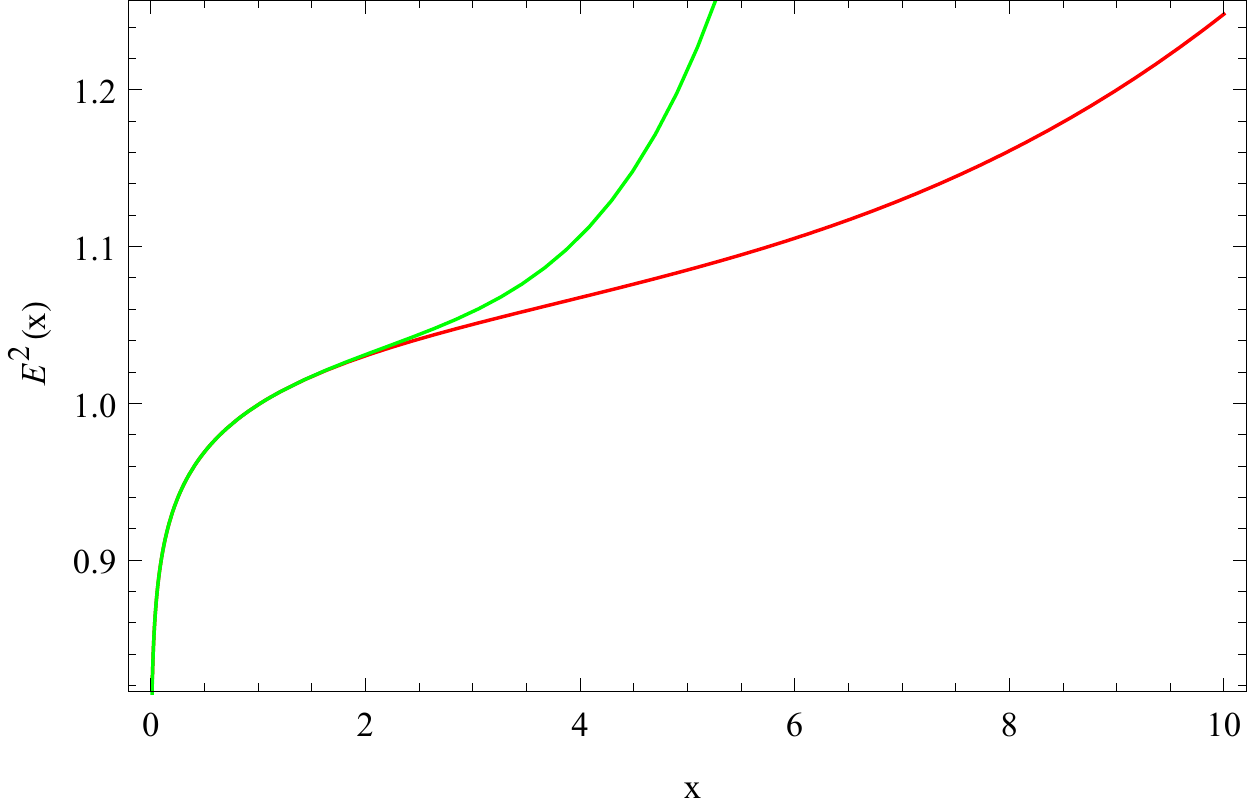}
\caption{\label{fig:i}Normalized Hubble parameter  $E^{2}(x)$ with redshift ($x$ for range $0$ to $10$). We assume $\eta=0.001$, $\varepsilon = 0.02$, $\alpha=1.2$ and $\beta=0.2 $ and $-0.2$ for  $\bar{\Omega}^{0}_{m}\simeq 0.274$ and $\bar{\Omega}^{0}_{\lambda}\simeq 0.725$. Green  and red curves refer  to $\beta=0.2$ and $ -0.2 $,  respectively.}
\end{figure}

In our scenario (perturbation + interaction) the $O_{m}(x)$-diagnostics is given as

\begin{eqnarray}
O_{m}(x)=\frac{Ax^{3\varepsilon /1+\alpha} + Bx^{4+3\varepsilon /1-\beta}+Cx^{\eta}-1}{x^{3}-1}\label{n31}
\end{eqnarray}
where \[A=(1+M)\bar{\Omega}_{\lambda}^{0},\] \[B=(1-N)\bar{\Omega}_{r}^{0}\] and \[C=\bar{\Omega}_{m}^{0}+N\bar{\Omega}_{r}^{0}-M\bar{\Omega}_{\lambda}^{0}.\]

\begin{figure}[h]
\centering  \begin{center} \end{center}
\includegraphics[width=0.50\textwidth,origin=c,angle=0]{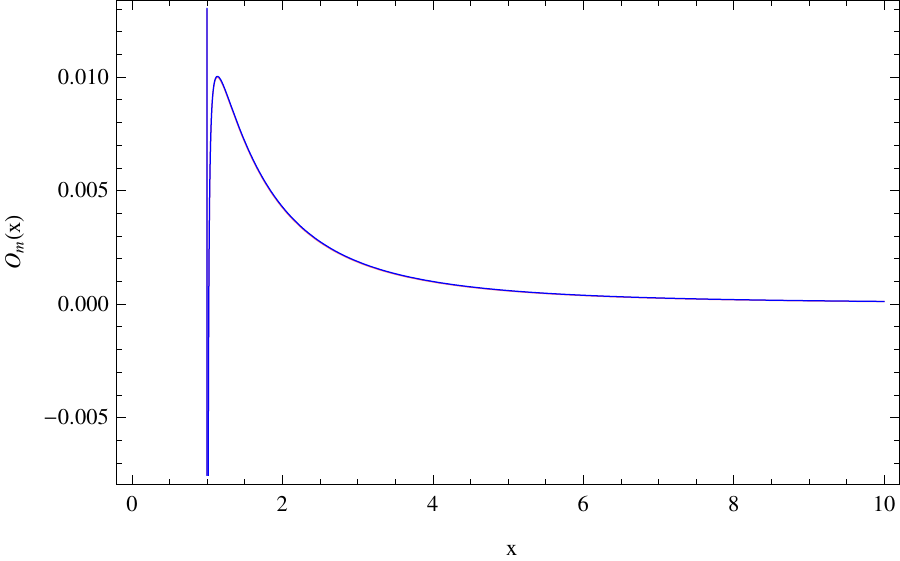}
\caption{\label{fig:i}Plot of $O_{m}(x)$-diagnostics with redshift ($x$ for range $0$ to $10$). We assume $\eta=0.001$, $\varepsilon = 0.02$, $\alpha=1.2$ and $\beta=1.2$ and $-1.2$ for $\bar{\Omega}^{0}_{\lambda}\simeq 0.725$, $\bar{\Omega}^{0}_{m}\simeq 0.274$ and $\bar{\Omega}^{0}_{r}\simeq 0.00005$. The plot is independent of the chosen values of $\beta$.}
\end{figure}

\begin{figure}[h]
\centering  \begin{center} \end{center}
\includegraphics[width=0.50\textwidth,origin=c,angle=0]{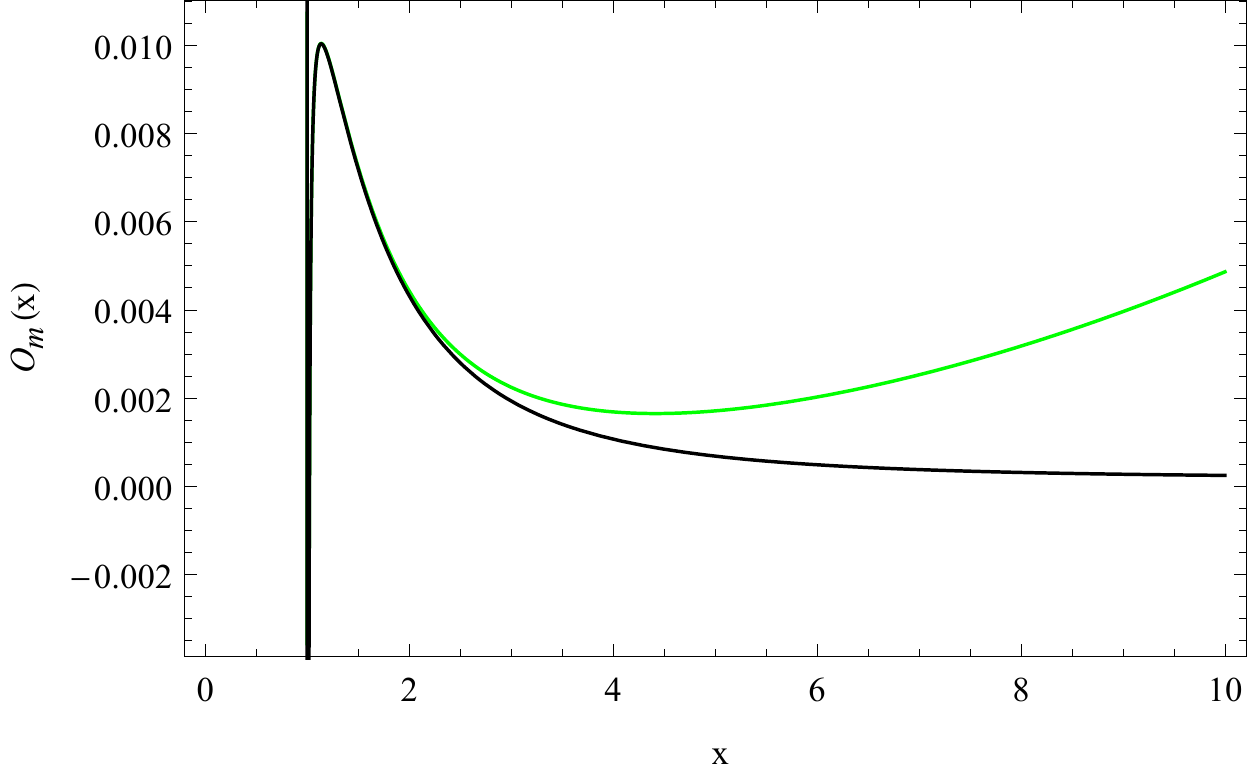}
\caption{\label{fig:i} Plot of $O_{m}(x)$-diagnostics with redshift ($x$ for range $0$ to $10$). We assume $\eta=0.001$, $\varepsilon = 0.02$, $\alpha=1.2$ and $\beta=-0.2$ and $-0.2$ for $\bar{\Omega}^{0}_{\lambda}\simeq 0.725$, $\bar{\Omega}^{0}_{m}\simeq 0.274$ and $\bar{\Omega}^{0}_{r}\simeq 0.00005$. Green and black curves  are plotted for $\beta=0.2$ and $-0.2$ respectively. }
\end{figure}


The difference of the squares of the normalized Hubble parameter $E^{2}(x)$ at two different redshifts $x_{i}$ and $x_{j}$ is given as
\begin{eqnarray}
\Delta E^{2}(x_{i},x_{j})=\bar{M}\bar{\Omega}^{0}_{\lambda}(x_{i}^{3\varepsilon/1+\alpha}-x_{j}^{3\varepsilon/1+\alpha}) +\nonumber \\ \bar{N}\bar{\Omega}^{0}_{r}(x_{i}^{4+3\varepsilon/1-\beta}-x_{j}^{4+3\varepsilon/1-\beta})+\chi(x_{i}^{\eta}-x_{j}^{\eta})\label{m1}
\end{eqnarray} where \[\Delta E^{2}(x_{i},x_{j})=E^{2}(x_{i})-E^{2}(x_{j}).\]

Measuring the values of $\Delta E^{2}(x_{i},x_{j})$ from observations we can estimate the proportionality constants.

\begin{table}[h]

\caption{First column shows the   Hubble parameter $H(z)$. Second and third columns  give the  redshifts  $z_{i}$ and  $x_{i}=1+z_{i}$  where $i=1,2,3,4$.}
\label{tab:1}

\begin{tabular}{lll}

\hline\noalign{\smallskip}
$H(z)$ & $z_{i}$ & $x_{i}$  \\

\noalign{\smallskip}\hline\noalign{\smallskip}
$69 \pm 12$ & 0.1 & 1.1 \\
$95 \pm 17$ & 0.4 & 1.4 \\
$168 \pm 17$ & 1.3 & 2.3 \\
$202 \pm 40$ & 1.75. & 2.75 \\
\noalign{\smallskip}\hline

\end{tabular}

\end{table}

We further  take a set of values  $H(z)$  at different redshifts given  in Table 1 \cite{c1}.  From these  values,   we obtain  six non-linear equations. Thus,  from (\ref{m1}) with  $H_{0}=73.8\pm2.4 \approx 73.8$  \cite{c2} we obtain  six values of   $\Delta E^{2}(x_{i},x_{j})$  mentioned  in  Table 2.

\begin{table}[h]

\caption{First column gives the  difference of squared normalized Hubble parameter $\Delta E^{2}(x_{i},x_{j})$, Second column is its numeric value and  third column is pairs of redshifts, where $i=1,2,3,4$ and $j=1,2,3,4$. }
\label{tab:1}

\begin{tabular}{lll}

\hline\noalign{\smallskip}
$\Delta E^{2}(x_{i},x_{j})$ & Numeric Value  & $(x_{i}, x_{j})$  \\

\noalign{\smallskip}\hline\noalign{\smallskip}
$\Delta E^{2}(x_{1},x_{2})$ & $-0.783$  & $(1.1, 1.4)$ \\
$\Delta E^{2}(x_{1},x_{3})$ & $-4.308$  & $(1.1, 2.3)$ \\
$\Delta E^{2}(x_{1},x_{4})$ & $-6.618$  & $(1.1, 2.75)$ \\
$\Delta E^{2}(x_{2},x_{3})$ & $-3.525$  & $(1.4, 2.3)$ \\
$\Delta E^{2}(x_{2},x_{4})$ & $-5.835$  & $(1.4, 2.75)$ \\
$\Delta E^{2}(x_{3},x_{4})$ & $-2.310$  & $(2.3, 2.75)$ \\
\noalign{\smallskip}\hline

\end{tabular}

\end{table}

With the help of these two Tables  we have following six non-linear equations for $\Delta E^{2}(x_{i},x_{j})$

\begin{eqnarray}
    A_{1}[(1.1)^{3\varepsilon /1+\alpha}-(1.4)^{3\varepsilon /1+\alpha}]+ \nonumber \\ A_{2}[(1.1)^{4+3\varepsilon /1-\beta}-(1.4)^{4+3\varepsilon /1-\beta}] \nonumber \\  + C[(1.1)^{\eta}-(1.4)^{\eta}]=-0.783\label{m2}
\end{eqnarray}

\begin{eqnarray}
    A_{1}[(1.1)^{3\varepsilon /1+\alpha}-(2.3)^{3\varepsilon /1+\alpha}]+ \nonumber \\ A_{2}[(1.1)^{4+3\varepsilon /1-\beta}-(2.3)^{4+3\varepsilon /1-\beta}] \nonumber \\  + C[(1.1)^{\eta}-(2.3)^{\eta}]=-4.308\label{m3}
\end{eqnarray}

\begin{eqnarray}
    A_{1}[(1.1)^{3\varepsilon /1+\alpha}-(2.75)^{3\varepsilon /1+\alpha}]+ \nonumber \\ A_{2}[(1.1)^{4+3\varepsilon /1-\beta}-(2.75)^{4+3\varepsilon /1-\beta}] \nonumber \\  + C[(1.1)^{\eta}-(2.75)^{\eta}]=-6.618\label{m4}
\end{eqnarray}

\begin{eqnarray}
    A_{1}[(1.4)^{3\varepsilon /1+\alpha}-(2.3)^{3\varepsilon /1+\alpha}]+ \nonumber \\ A_{2}[(1.4)^{4+3\varepsilon /1-\beta}-(2.3)^{4+3\varepsilon /1-\beta}] \nonumber \\  + C[(1.4)^{\eta}-(2.3)^{\eta}]=-3.525\label{m5}
\end{eqnarray}

\begin{eqnarray}
    A_{1}[(1.4)^{3\varepsilon /1+\alpha}-(2.75)^{3\varepsilon /1+\alpha}]+ \nonumber \\ A_{2}[(1.4)^{4+3\varepsilon /1-\beta}-(2.75)^{4+3\varepsilon /1-\beta}] \nonumber \\  + C[(1.4)^{\eta}-(2.75)^{\eta}]=-5.835\label{m6}
\end{eqnarray}

\begin{eqnarray}
    A_{1}[(2.3)^{3\varepsilon /1+\alpha}-(2.75)^{3\varepsilon /1+\alpha}]+ \nonumber \\ A_{2}[(2.3)^{4+3\varepsilon /1-\beta}-(2.75)^{4+3\varepsilon /1-\beta}] \nonumber \\  + C[(2.3)^{\eta}-(2.75)^{\eta}]=-2.310\label{m7}
\end{eqnarray}

where \[A_{1}=\bar{M}\bar{\Omega}^{0}_{\lambda},\] \[A_{2}=\bar{N}\bar{\Omega}^{0}_{r}.\]

\section{$O_{m}3(x)$ diagnostics and Statefinders}
\label{sec:4}
The two point $O_{m}(x)$-diagnostics may be defined as\cite{a7}

\begin{eqnarray}
O_{m}(x_2,x_1)=\frac{E^{2}(x_{2})-E^{2}(x_{1})}{x_{2}^{3}-x_{1}^{3}}\label{n32}
\end{eqnarray}
and $3$-point $O_{m}(x)$-diagnostics as

\begin{eqnarray}
O_{m}(x_3,x_2,x_1)=\frac{O_{m}(x_2,x_1)}{O_{m}(x_3,x_1)}.\label{n33}
\end{eqnarray}

Using (\ref{n32}) and (\ref{n33}) we have $O_{m}3(x)$-diagnostic

\begin{eqnarray}
O_{m}(x_3,x_2,x_1)=\frac{E^{2}(x_{2})-E^{2}(x_{1})}{x_{2}^{3}-x_{1}^{3}}.\frac{x_{3}^{3}-x_{1}^{3}}{E^{2}(x_{3})-E^{2}(x_{1})}\label{n34}
\end{eqnarray}

$E^{2}(x_{3})$,  $E^{2}(x_{2})$ and $E^{2}(x_{1})$ can be calculated from (\ref{n28}) at three different redshifts. We would take up this study in future. Here, we use the statefinders for constraining the interaction.

  In  addition to the cosmological constant  several  other  candidates for  dark energy (quintom, quintessence, brane-world, modified gravity etc., e.g. \cite{a8}) have been  proposed. To differentiate different models of dark energy Sahni \emph{et.al} \cite{a9} proposed statefinder   diagnostics  based on dimensionless parameters $(r, s)$  which are the functions  of scale factor and its time derivative. These parameters are define as

\[r=\frac{\dddot{a}}{aH^{3}}\] and \[s=\frac{r-1}{3(q-\frac{1}{2})}\] with deceleration parameter \[q=-\frac{\ddot{a}}{aH^{2}}\]

Thus, re-writng the statefinders as,

\begin{eqnarray}
r=\frac{\ddot{H}}{H^{3}}+\frac{3\dot{H}}{H^{2}}+1\label{n35}
\end{eqnarray}

\begin{eqnarray}s=-\frac{2}{3(3H^{2}+2\dot{H})}\left[\frac{\ddot{H}}{H}+3\dot{H}\right] \label{n36}.\end{eqnarray}

Considering for interaction + perturbation scenario the statefinder parameters may be  calculated as from (\ref{n28}).
Let
\begin{eqnarray}H^{2}(x)=H^{2}_{0}y_{0}\label{n37}\end{eqnarray}
where
\begin{eqnarray} y_{0}=Ax^{3\varepsilon/1+\alpha}+ B x^{4+3\varepsilon/1-\beta}+ C x^{\eta}\label{n38}\end{eqnarray}

\begin{eqnarray}
r=\frac{y_{2}\dot{x}^{2}}{2H^{2}_{0}y^{2}_{0}}+\frac{y_{1}\dot{x}\ddot{x}}{2H^{2}_{0}y^{2}_{0}}-\frac{y^{2}_{1}\dot{x}^{2}}{2H^{2}_{0}y^{3}_{0}}(1-3H^{2}_{0}y_{0})+1\label{n39}
\end{eqnarray}

where

\begin{eqnarray} y_{1}=A\left(\frac{3\varepsilon}{1+\alpha}\right)x^{\frac{3\varepsilon}{1+\alpha}-1}\nonumber\\+B\left(\frac{4+3\varepsilon}{1-\beta}\right)x^{\frac{4+3\varepsilon}{1-\beta}-1}\nonumber\\+C\eta x^{\eta-1}\label{n40} \end{eqnarray}

\begin{eqnarray} y_{2}=A\left(\frac{3\varepsilon}{1+\alpha}\right)\left(\frac{3\varepsilon-\alpha-1}{1+\alpha}\right)x^{\frac{3\varepsilon}{1+\alpha}-2}\nonumber\\+B\left(\frac{4+3\varepsilon}{1-\beta}\right)\left(\frac{3+3\varepsilon+\beta}{1-\beta}\right)x^{\frac{4+3\varepsilon}{1-\beta}-2}\nonumber\\+C\eta(\eta-1) x^{\eta-2}\label{n41} \end{eqnarray}

From (\ref{n39}),(\ref{n40}) and (\ref{n41}) we can find the other statefinder parameter $s$  defined  by (\ref{n36}) as given below

\begin{eqnarray}s=\zeta\left[\frac{H_{0}y_{2}\dot{x}^{2}}{y^{1/2}_{0}}+\frac{H_{0}y_{1}\dot{x}\ddot{x}}{y^{1/2}_{0}}-\frac{y_{1}\dot{x}}{y_{0}}+3H^{2}_{0}y_{1}\dot{x}\right] \label{n42} \end{eqnarray}
where $\zeta$ given as

\begin{eqnarray}\zeta=-\frac{1}{3(3H^{3}_{0}y^{3/2}_{0}+2H^{2}_{0}y_{1}\dot{x})}\label{n43} .\end{eqnarray}

For non-interacting model i.e., $\alpha=0$ and $\beta=0$  and in the absence of perturbation these turn out to  be the  same as in $\Lambda$CDM model $(r=1, s=0)$ for $(w_{\lambda}=-1)$ and $(w_{m}=0)$ with negligible contribution of radiation.

\section{Discussion}
\label{sec:6}
 We argue that the contributions from the recent acceleration of the universe must generate the tensor fluctuations in a way similar to early inflation. It would amplify the  tensor-to-scalar ratio. In fact, being robust and of recent and much longer duration this may explain the high value of $r$  recently observed by BICEP2, since it must represent  a the sum of inflationary as well as well as the post-inflationary current acceleration. We have used   the  single tachyonic scalar field which splits due to some  unknown (apparently Higgs-like) mechanism into three components (cosmological constant, radiation and dust matter) which interact to achieve a thermodynamical equilibrium. The interaction among the components  at present may boost  the tensor-to-scalar amplitude ratio caused by the early inflation and must appear as the enhanced signature beyond the bounds set by the Planck observations.  The entire evolution of the universe arises from this process of interaction.  Due to consideration of radiation in this field  the dust matter appears with negative energy.  A small perturbation allowed in EoS of cosmological constant changes its status from a true cosmological constant to a shifted cosmological parameter.  Similarly,  status of radiation and dust matter changes to shifted radiation and shifted exotic matter. With the perturbation shifted exotic matter gains non-zero negative pressure which also helps (along with dark energy)  in the accelerated expansion of the universe. Total energy of field stays conserved but the field components mutually interact with interaction strength parameter $Q$ resulting in local violation of energy conservation. The interaction reflects in diagnostics with our choice of interaction strengths  which are  constrained by the astrophysical data of Hubble parameter at different redshifts chosen ( $z=0.1, 0.4, 1.3$ and $1.75$) (albeit a narrower  range of redshifts  would   be  more useful in determining dynamics at an epoch).  The  $O_{m}3(x)-$ diagnostics and Statefinder parameters  may be further explored  for   interaction $+$ perturbation model to check the important constraints.

\begin{acknowledgments}
 The authors thank the   University Grants Commission, New Delhi, India for support under major research project F. No. 37-431/2009 (SR).
\end{acknowledgments}

\end{document}